\documentclass[final, 3p]{elsarticle}
\usepackage[utf8]{inputenc}
\usepackage{url}
\usepackage[normalem]{ulem}
\usepackage{parskip}

\usepackage{amsmath}
\usepackage{physics}
\usepackage{caption}
\usepackage{subcaption}

\usepackage{todonotes}

\usepackage{soul, xcolor} 

\setstcolor{teal} 

\usepackage[english]{babel}

\usepackage{amstext}
\usepackage{amssymb}
\usepackage{graphicx}
\usepackage{adjustbox}
\graphicspath{{./Images/}}
\usepackage{dsfont}

\usepackage{textcomp}
\usepackage{listings}
\usepackage{relsize}
\usepackage{multicol}
\usepackage{hyperref}
\hypersetup{
	colorlinks=true,
	linkcolor=black,
	filecolor=black,      
	urlcolor=black,
	citecolor=blue,
}

\usepackage{physics}
\usepackage{url}

\usepackage{algorithm}
\usepackage{algpseudocode}
\usepackage{pifont}

\usepackage{bbm}
\usepackage{mathtools}
\usepackage{hhline}
\usepackage{tabularx}

\let\today\relax
\makeatletter
\def\ps@pprintTitle{%
    \let\@oddhead\@empty
    \let\@evenhead\@empty
    \def\@oddfoot{\footnotesize\itshape
         {Preprint} \hfill\today}%
    \let\@evenfoot\@oddfoot
    }
\makeatother
\begin{document}

\begin{frontmatter}
\title{Backcasting COVID-19:\\ A Physics-Informed Estimate for Early Case Incidence}

\author[jhu,umass]{G.A. Kevrekidis}
\author[illinois]{Z. Rapti}
\author[JRC]{Y. Drossinos}
\author[umass]{P.G. Kevrekidis}
\author[umass]{M.A. Barmann}
\author[umass]{Q.Y. Chen}
\author[seville1,seville2]{J. Cuevas-Maraver}

\address[jhu]{Department of Applied Mathematics and Statistics, Johns Hopkins University, Baltimore, MD 21218, USA}

\address[illinois]{Department of Mathematics and Carl R.Woese Institute for Genomic Biology, University of Illinois
at Urbana-Champaign}

\address[JRC]{European Commission, Joint ResearchCentre, I-21027 Ispra
  (VA), Italy}

\address[umass]{Department of Mathematics and Statistics, University of Massachusetts Amherst,
Amherst, MA 01003, USA}

\address[seville1]{Grupo de F\'{\i}sica No Lineal, Departamento de F\'{\i}sica Aplicada I,
Universidad de Sevilla. Escuela Polit\'{e}cnica Superior, C/ Virgen de \'{A}frica, 7, 41011-Sevilla, Spain}

\address[seville2]{Instituto de Matem\'{a}ticas de la Universidad de Sevilla (IMUS). Edificio
Celestino Mutis. Avda. Reina Mercedes s/n, 41012-Sevilla, Spain}

\begin{abstract}
It is widely accepted that the number of reported cases during the first stages of the 
COVID-19 pandemic severely underestimates the number of actual cases. 
We leverage delay embedding theorems of Whitney and Takens and 
use Gaussian Process regression to estimate the number
of cases during the first 2020 wave based on the second wave of the epidemic in
several European countries, South Korea, and Brazil. We assume that the second wave was more accurately monitored and hence that it can be trusted. We then construct a manifold diffeomorphic to that of the implied original dynamical system, using fatalities or hospitalizations only. Finally, we restrict the diffeomorphism to the reported cases coordinate of the dynamical system. Our main finding is that in the European countries studied, the actual cases are under-reported by as much as 50\%. On the other hand, in South Korea - which had an exemplary and proactive mitigation approach - a far smaller discrepancy between the actual and reported cases is predicted, with an approximately 17\% predicted under-estimation. We believe that our backcasting framework is applicable to other epidemic outbreaks where (due to limited or poor quality data) there is uncertainty around the actual cases.

\end{abstract}


\end{frontmatter}

\section{Introduction}
During the early stages of COVID-19 (and in fact of any emerging disease) even estimating the basic reproduction number $R_0$ was a challenge \cite{siettos2020, kolokolnikov2021, omori2020, viceconte2020} due to lack of information and the absence or poor quality of data. $R_0$ is defined as 
the number
of secondary infections an infectious individual can generate in a population of susceptible individuals. It is quite important for
epidemiologists and public health officials to have an accurate
estimate of its magnitude.
The clinical characteristics of the disease - latency period, period of infectiousness, incubation period - were not known, so differing estimates of $R_0$ existed (ranging from 1.4-6.49) \cite{liu2020} that largely depended on the models that were used to estimate them and the corresponding assumptions made. At the same time, limited testing capacity obscured the true size of the epidemic and the actual growth rate.
Varying  estimates of symptomatic and
asymptomatic cases also exist \cite{ing2020, bhatta2020, sah2021}.
These inadequacies, in turn, caused serious handicaps in mounting an appropriate response by public health authorities. While our understanding of the relevant models and both their benefits and weaknesses has greatly progressed \cite{GNANVI2021258,holmdahl2020,doi:10.1177/0272989X21990391}, 
there is still significant room for improvement of our understanding of both the first and subsequent waves of the epidemic.

As the epidemic progresses and more data become available - daily new cases, number of tests performed, daily new hospital admissions, daily new deaths, etc. - these too should be taken with a grain of salt. It is now widely known, and generally accepted, that COVID-19 reported data are neither reliable nor complete, and that the best practice is to base mathematical models on hospitalization and fatality time series data \cite{holmdahl2020}. Reported-case count time series are unreliable due to the limited number of tests available initially, and the large number of cases that are asymptomatic \cite{bhatta2020} and/or go unreported \cite{li2020}. A recent systematic review and data meta-analysis concluded that 35.1\% [95\% CI: 30.7 - 39.9\%] were asymptomatic infections \cite{sah2021}.

It is then natural that various methods and studies exist attempting to reconstruct the true number of case counts and fatalities. In \cite{phippsl2020} a backcasting approach based on fatalities and a Gamma distribution of the time from infection to death was used to study the epidemic in 15 countries. It was estimated that the number of infected people is 6.2 [$95 \% $ CI: $4.3-10.9$] times higher than reported. This echoes a study of the epidemic in the US \cite{wu2020}, where probabilistic bias analysis was used to approximate the true case counts. They found that $89\%$ of infections were undocumented.
Indeed, the US Centers for Disease Control indicated an under-reporting of infections by a factor of 2 to 13 times in \cite{cdc}, illustrating the gravity and relevance of the issue. In Brazil, it was estimated \cite{paixao2021} that the actual case counts are three times and the deaths twice as many as those reported.

A study of the epidemic dynamics around the globe \cite{russell2020} using a Bayesian Gaussian Process model found large disparities in the degree of under-reporting among countries. Cumulative infection data from several European countries were studied using the so-called capture-recapture methods \cite{bohning2020} and it was found that the ratio of calculated total over observed (i.e., reported) cases was around 2.3. In \cite{cuevas2021} we chose to neglect infection data and instead focused on fatalities, for a study of the epidemic in Mexico. Fitting the model to the infection time series produced unreliable predictions of future deaths, due to the under-reporting of infections.

While fatalities are believed to be more reliable than case counts~\cite{cuevas2021}, care should still be taken when using them as a benchmark due to different ways of measuring and reporting the data among countries \cite{beaney2020}. For instance, excess deaths, rather than reported deaths, during the first wave in the northern part of Italy were embedded in a differences-in-differences regression model. One of the findings of the study was that deaths may be under-reported by as much as $60\%$ \cite{ciminelli2020}. 
The reason for the discrepancy is that only hospital deaths are included in the official reports.
The excess mortality in the first months of the epidemic was estimated to be $28\%$ higher than the reported COVID-19 fatalities. Another pertinent point is that delays in death counts may be as long as a year in some cases \cite{weinberger2020}.  This limits our ability to produce accurate, real-time forecasts if recent data are not reliable; but if the vast majority of the data is eventually counted and correctly revised, then this situation lends well to a retrospective analysis after, say, a year.

In the present study, we develop and apply a method that can be used to extrapolate from the second (and generally latter) wave, the fatalities and the COVID-19 infections. We assume that during the second wave both reported infections and reported fatalities are accurate and are used together with the fatalities during the first wave (also assumed to be accurate) to backcast the reported infections during the first wave. 
This method can be applied to countries whose COVID-19 time series are characterized by a first peak, succeeded by a period of very low daily incidence during the summer months (i.e., between the first and the second wave), followed by a second peak in the fall. Using the second-wave time series as training sets for our algorithm, we carry out this program for a substantial number of countries, principally from Europe, but also from Asia and America, obtaining a consistent under-reporting of the relevant diagnostics. 

Our presentation is structured as follows. First we provide the mathematical background for our analysis in Section \ref{sec:Theory}. Subsequently, we present the relevant data and their structure in Section \ref{sec:Data},
comment on uncertainty in diagnosed cases in \ref{sec:Data Uncertainty}, and we present a schematic of our
methodology in \ref{sec:backcasting algorithm}. In Section \ref{sec:Numerical Results} we present our numerical findings, and Section \ref{sec:Discussion and Conclusion} wraps up the manuscript with some conclusions and directions for future work. The Supplementary Material contains further information on our methodology, describes model parameters, and addresses issues concerning data uncertainty and hyperparameter tuning.

\section{Method}

\subsection{Theory: Delay embedding}
\label{sec:Theory}
Our operating assumption is that we are given the time series of \textit{Reported Cases} ($C$) and \textit{Reported Fatalities (Deaths)} ($D$) per day in a geographical region over time. More specifically, we denote
\begin{align}
    D &= \qty{d_t}_{t\geq 0}^T\\
    C &= \qty{c_t}_{t\geq 0}^T
\end{align}
where $d_t, c_t$ denote the fatalities and  officially diagnosed cases on day $t$, for $t = 0,1,2,\ldots,T$. We further assume that $D$ is \textit{accurate} for all $t$ while $C$ is only accurate after time $t\geq T^*$ and \textit{inaccurate} for $t<T^*$, where $T^*$ is a time point chosen between the first and second waves, chosen as explained in the following section.

Suppose that $d_t$ and $c_t$ are discrete observations of two corresponding quantities $D(t)$, $C(t)$ that satisfy an $n$-dimensional system of coupled deterministic ordinary differential equations whose dynamics feature a unique attractor. Then, the state of the system at any given time can be represented by an $n$-dimensional vector and the overall behavior of the system can be represented on a manifold $\mathcal{M}\subset\mathbb{R}^n$.

According to the delay embedding theorems of Whitney and Takens \cite{deyle2011, noakes1991, sauer1991, takens1981}, we are able to use only $d_t$ to construct a manifold $\mathcal{M}_D\subset\mathbb{R}^{k}$ that is diffeomorphic to $\mathcal{M}$. Such a manifold
is guaranteed to exist whenever the embedding dimension $k$ satisfies $k\geq 2n+1$: that is, there exists a differentiable invertible transformation (diffeomorphism) $f:\mathcal{M}_D\to\mathcal{M}$. Since $c_t$ is known after time $T^*$ we can use that part of the data set to \textit{learn} (using regression) the restriction of $f$ to the ``reported-cases" coordinate

\begin{equation}
 \hat{f}\sim f\big\vert_C:\mathcal{M}_D\to C(t), \qquad t \geq T^*
\end{equation}

We proceed to evaluate $\hat{f}$ on the points of the reconstructed manifold $\mathcal{M}_D$ observed for $t\leq T^*$ in order to estimate $c_t$ during the early stages of the pandemic.

\subsection{Data}
\label{sec:Data}
In Supplementary Material (SM), Appendix A, we summarize the sources of our data and we comment on their quality. In addition, we address issues of data uncertainty, we comment on the reliability of the time series used and on
the embedding and Gaussian Process parameters.

\subsubsection{Structure and $T^*$}
\label{sec:structure_and_t_star}
The framework proposed in Section \ref{sec:Theory} is very general in that it only requires two time series of the observed quantities ($C, D$) and some knowledge of the time $T^*$.

$T^*$ is the time after which $C$ can be considered \textit{accurate}, and it may be hard to define. The naive definition of accurate and inaccurate is \textit{true} and \textit{false} respectively, meaning that we take all information after time $T^*$ at face value. It is arguably more natural to think of the \textit{accuracy} of the considered data as a continuously varying quantity as opposed to a Boolean variable. In principle, it may be possible to incorporate additional statistics such as (but not limited to) the amount of testing \cite{hasell2020} and vaccination rates \cite{mathieu} to quantify the uncertainty of the data. However, this process would complicate our analysis since the availability, quality, and adequacy of additional markers is not certain, and their consideration would result in a more complex but less applicable model.

\begin{figure}[h]
  \begin{subfigure}[b]{0.49\textwidth}
    \includegraphics[width=\textwidth]{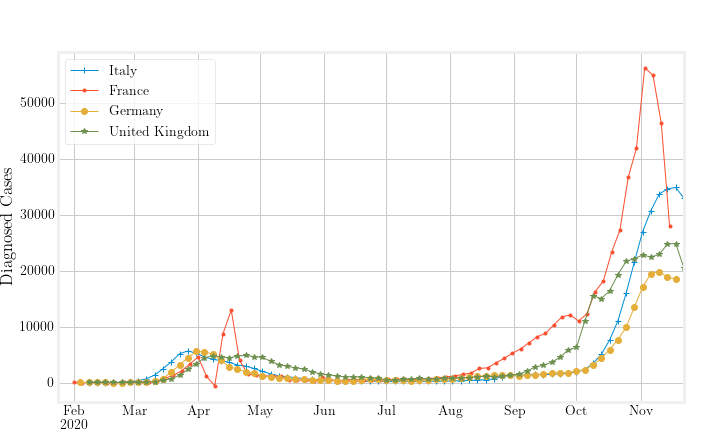}
    \caption{}
    \label{fig:structure_of_cases}
  \end{subfigure}
  \hfill
  \begin{subfigure}[b]{0.49\textwidth}
    \includegraphics[width=\textwidth]{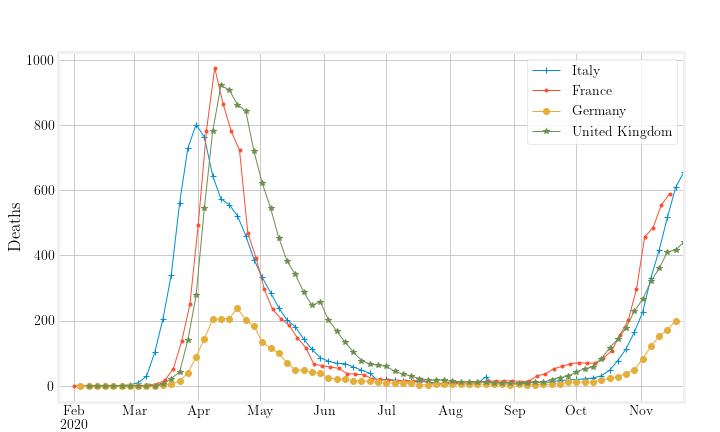}
    \caption{}
    \label{fig:structure_of_deaths}
  \end{subfigure}
  \caption{Seasonal structure of the (a) case and (b) death time series for the UK and the three most populous countries in the EU (Germany, France, Italy) during the first months of 2020. 
  (The UK officially exited the EU after the 31st of Janyary, 2020). Few diagnosed cases and deaths were observed during the summer months.}
\end{figure}

Fortunately, there is structure in many time series at the country and local level that can be exploited. Several countries in Europe experienced a first peak in the observed cases and fatalities,
as shown in Figures \ref{fig:structure_of_cases} and \ref{fig:structure_of_deaths}, respectively.
This was followed by a decline during the summer months, before the second peak in the fall. 
Thus, a heuristic definition of $T^*$ can be described as the time when mortality falls close to zero for a long period, which coincides, in many countries, with the summer months of 2020. The  leads to an \textit{almost boolean} $T^*$ occurring sometime within this period.

We note that our backcasting results are not sensitive to such a $T^*$, as long as its choice is reasonable, namely falls within the period of low mortality and not close to the second peak
(see Section \ref{Italy_case_study} for the effect of the choice of $T^*$ on backcasting results for Italy). A compromise in the characterization of \textit{accuracy} can be reached by assuming that the data after $T^*$ are not necessarily true, but can be trusted enough to make \textit{an estimate} of the true number of infections after $T^*$, which is further explored in Section \ref{sec:uncertainty_in_cases}.

An additional complication arises due to varying mortality rates as the epidemic progressed. For instance, a study for Sweden \cite{stralin2021} spanning data from March  to September 2020 found that mortality in hospitalized patients decreased as the pandemic progressed. This may be attributed to improvements in treatment protocols, the availability of new treatments [e.g. remdesivir, monocolonal antibodies], differences in the virulence of circulating variants \cite{lan2021}, or \text{differing} age distribution of those infected in the first versus second wave \cite{james2021}. 
The inability to address these issues fully is why our results are qualified as approximate estimates of the true case incidence.

\subsection{Uncertainty in Diagnosed Cases}
\label{sec:Data Uncertainty}

\label{sec:uncertainty_in_cases}
We are interested in building an uncertainty estimate around the diagnosed cases we use during our calibration period, which will, in turn, correspond to uncertainty around the backcasted result. We turn to the literature to obtain a ``hidden-case" estimator that preserves some of the integrity of our current data. 

The Capture-Recapture (CRC) method presented in \cite{bohning2020} estimates that if the diagnosed cases at time $t$ are $c_t$, then the true number of cases at time $t$ is a random variable $\hat{c}_t$, where~\cite{bohning2020}, \cite{zhao2020preliminary}:
\begin{align}
    \mathbb{E}\qty[\hat{c}_t] &= c_t + \frac{(c_t)^2}{c_{t-1}-d_t}\\
    \mathbb{V}\qty[\hat{c}_t] &= \frac{(c_t)^4}{(1+c_{t-1}-d_t)^3}+\frac{4(c_t)^3}{(1+c_{t-1}-d_t)^2}+\frac{(c_t)^2}{1+c_{t-1}-d_t}.
\end{align}

Thus, the case time series after $T^*$ can be adjusted (accompanied by a $95\%$ confidence interval) using case and fatality observations after $T^*$. This estimator only depends on our available time-series information, and does not take into account any external prior knowledge (such as serial interval distribution, etc). This is desirable because we believe that post-$T^*$ diagnosed case counts are of much higher fidelity than early ones, and we would like our estimates to be based on them, since they are true observables of the system behavior we are trying to capture.

Our results show that this estimate cannot be used reliably in this problem framework \textit{without} backcasting, since $\hat{c}_t$ still depends on the validity of $c_t$, which is not close to the true number of cases before $T^*$. We choose to present the main part of our numerical results using CRC to quantify uncertainty, aside from Figures \ref{fig:Italy_BC_Daily} and \ref{fig:Italy_BC_Cumulative} in Section~\ref{Italy_case_study} where we also present the incorporated Gaussian-Process (GP) uncertainty estimates. The uncertainty estimation method, however, is arbitrary, and one can choose it independently as part of the backcasting algorithm.

\subsection{The Backcasting Algorithm}
\label{sec:backcasting algorithm}
\begin{figure}[h]
    \includegraphics[width=\textwidth]{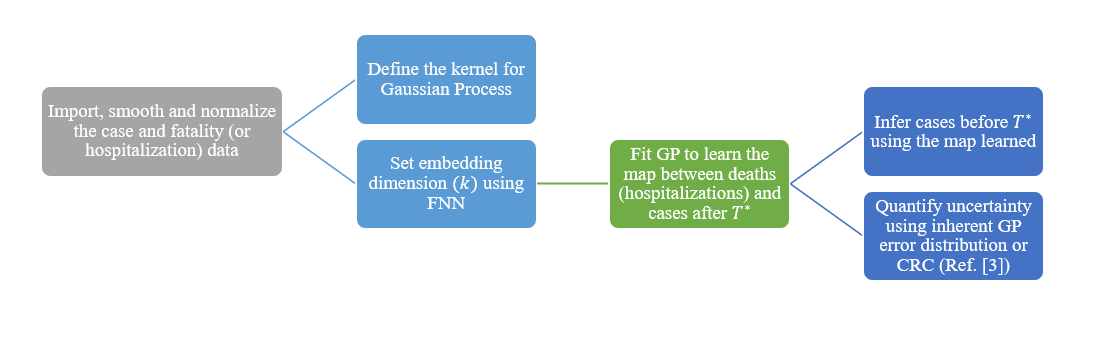}
    \caption{Flowchart of the proposed Backcasting Algorithm. The acronyms are: FNN: False-Nearest-Neighbors; GP: Gaussian Process (regression); CRC: Capture-Recapture (method).}
    \label{fig:flowchart}
\end{figure}

Figure~\ref{fig:flowchart} summarizes the algorithm we used to backcast the number of cases, given two time
series. We chose the time series of reported number of cases and number of fatalities. First, the embedding
dimension $k$ was determined via the False-Nearest-Neighbors (FNN) \cite{abarbanel1993} (or alternatively Average False Neighbors (AFN) \cite{cao1997afn}) algorithm as implemented in the 
Python nonlinear time series module \texttt{nolitsa}\footnote{https://github.com/manu-mannattil/nolitsa} \cite{Mannattil_2016_nolitsa}, \cite{mannattil2017applicability_nolitsa}. Then, we used
Gaussian Process (GP) regression (see SM, Appendix B)
to obtain $\hat{f}_\text{GP}$, a process that requires learning the case time series after time $T^*$ and then using it to backcast the number of case before $T^*$. We used the \texttt{sklearn}\footnote{https://scikit-learn.org/stable/modules/gaussian\_process.html} Python library \cite{scikit-learn} to perform the Gaussian Process regression. Lastly, we quantified uncertainty estimates using the capture-recapture method of Ref.~\cite{bohning2020}. We note in passing that further work on the use of GP for learning dynamical system behavior in the same manner can be found in~\cite{lee2019linking}. 

Importantly, each step of the proposed algorithm is independent. One may use GP if other reliable uncertainty estimates are not available, but may also use another regression method to learn the estimated function. Similarly one may also use a separate uncertainty method specific to the country or data set the algorithm is applied to. This is impractical when working with multiple countries, but it will lead to more accurate results when optimizing these choices for a specific data set. The theory behind the validity of the algorithm is valid regardless of the method applied.

In SM, Appendix A.3, we provide further information on the definition and use of the embedding parameters (the delay time $\tau$ and
the embedding dimension $k$) along with Gaussian Process parameters and the corresponding kernels used.

\section{Numerical Results}
\label{sec:Numerical Results}

\subsection{Italy: A Case Study}
\label{Italy_case_study}
We analyze the time series of a single country as a prototypical case example: Italy.

\begin{figure}[h!]
    \centering
    \includegraphics[width=13cm]{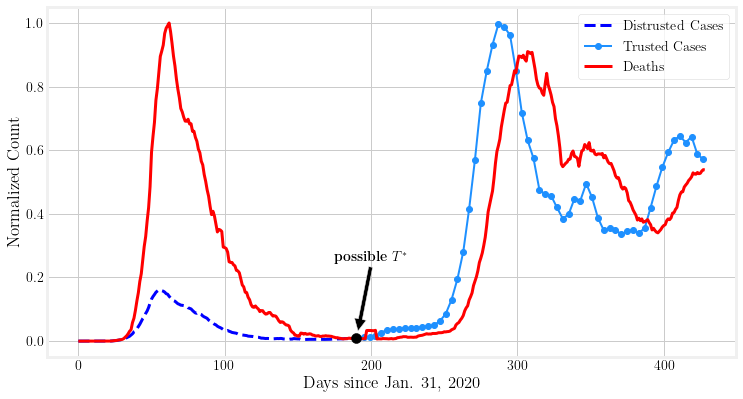}
    \caption{Normalized diagnosed cases and deaths in Italy. We divide each series by its maximum to be able to compare both quantities in one graph (that is $c_t/\max_t\qty{c_t},\quad d_t/\max_t\qty{d_t}$, where $\max_t\qty{c_t}=35073,\quad \max_t\qty{d_t}=814$).}
    \label{fig:Italy_case_death_comparison}
\end{figure}

In Figure \ref{fig:Italy_case_death_comparison} we present the normalized time series for Italy from the beginning of the pandemic. While the fatalities in the two \textit{waves} beginning at approximately $t=30$ and $t=250$, respectively, are of the same magnitude, we note that the diagnosed cases during the first wave are much lower than the second wave would suggest. In this case, a nominal $T^*$ is identified, but it could be specified to be at any point where the incidence of both cases and fatalities is very low (i.e., during the summer months). 

\begin{figure}[h!]
    \centering
    \includegraphics[width=7cm]{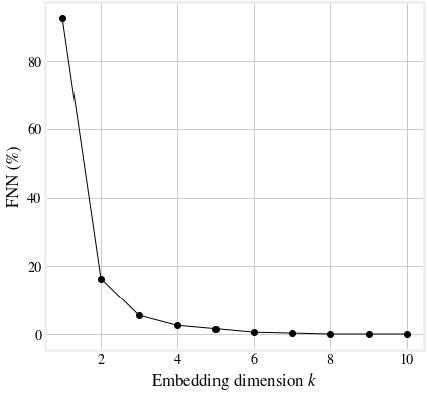}
    \caption{Number (in percentage) of False Nearest Neighbors for the death time series of Italy under the Chebyshev metric, ({see Figure} \ref{fig:Italy_case_death_comparison}). The algorithm is implemented through the Python module \texttt{nolitsa}.}
    \label{fig:FNN_for_Italy}
\end{figure}

Because the disparity between the mortality rates of the two waves suggested by Figure \ref{fig:Italy_case_death_comparison} is not trustworthy, and since the data satisfy the structure presented in Section \ref{sec:structure_and_t_star}, we claim that this is a good candidate time series to backcast. 
Using the FNN algorithm we estimate the embedding dimension for the fatality time series to be $k=10$ (see Figure~\ref{fig:FNN_for_Italy} where the percentage of false neighbors drops to $0$ at that dimension). In addition, we arbitrarily, {but consistently, as discussed,} pick $T^*=160$ occurring during the period of low case incidence. 

\begin{figure}[h!]
    \centering
    \includegraphics[width=14cm]{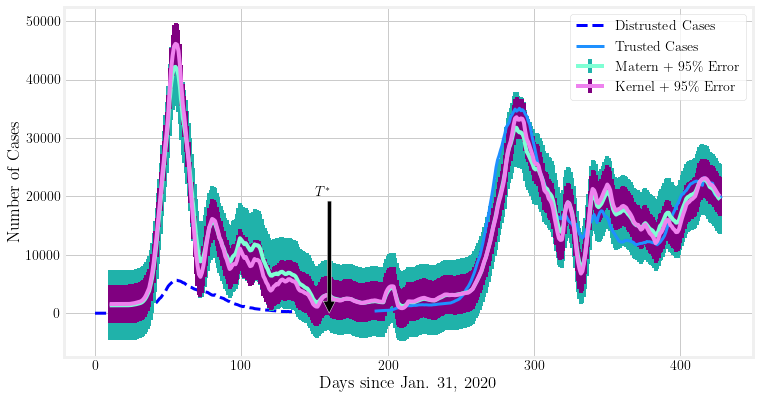}
    \caption{Backcasting result for Italy with a time window of 10 days. The panel includes results for two different choices of kernel (RBF and Mat\'{e}rn) for the GP regression.}
    \label{fig:Italy_BC_Daily}
\end{figure}

\begin{figure}[h!]
   \centering
    \includegraphics[width=14cm]{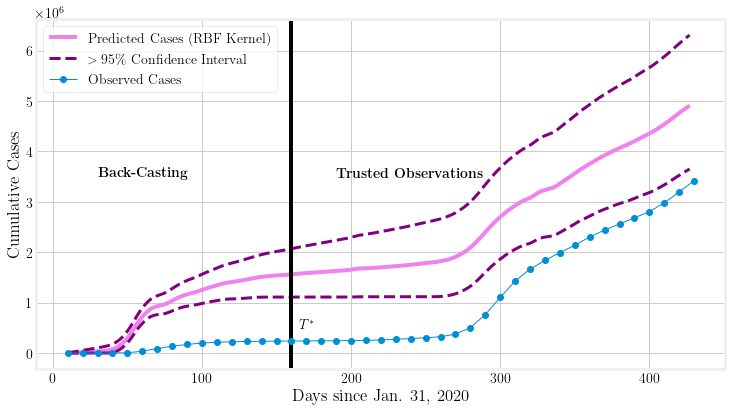}
    \caption{Cumulative backcasting result for Italy (with a time window of 10 days) corresponding to the daily results of Figure \ref{fig:Italy_BC_Daily}.
  The panel only presents one of the results (RBF) for the sake of clarity.  After time $T^*=160$, the predicted cases (pink) are almost an exact translation of the observed cases (blue) since the model has been trained at those points.}
    \label{fig:Italy_BC_Cumulative}
\end{figure}

Given {the embedding dimension} we construct $\mathcal{M}_{{D}}$ such that 
$\vb{d}_t=\qty{d_{t-9},...,d_{t}}$  $ \in \mathcal{M}_{D}$, proceeding to fit our GP estimator:
\begin{align}
    \hat{f}_\text{GP}(\vb{d}_t)=c_t,\qquad\forall t\geq 160
\end{align}
and then backcast to the initial cases $\hat{c}_t$:
\begin{align}
    \hat{c}_t = \hat{f}_\text{GP}(\vb{d}_t),\qquad\forall t\in[10,160]
\end{align}
We obtain the result shown in 
Figs.~\ref{fig:Italy_BC_Daily}-\ref{fig:Italy_BC_Cumulative}. A remarkable feature
of Fig.~\ref{fig:Italy_BC_Daily} is that at the peak of the first wave,
the number of reported daily cases was found to be
less than 10,000, while any selection of
the kernel in our backcasting scenarios
suggests that there were approximately 40,000 cases
at that time. Even adjusting for the trusted observations in Fig.~\ref{fig:Italy_BC_Cumulative}, we still find a substantial disparity between the number
of cumulative cases observed and the ones  predicted by our backcasting analysis.
The uncertainty interval of this and similar figures is computed based on the uncertainty of the corresponding daily cases, and is $>95\%$ based on the sub-additivity of standard deviation. As we consider more data, the disparity (as a ratio) between observed and estimated cases will necessarily decrease because the result is cumulative and the increments agree after $T^*$. Comparing cumulative cases at $T^*$ gives a more accurate picture of the beginning stages of the pandemic, while comparing at a later time can be used to understand the long-term effect of that early (unreported) spike in cases;
see, for example, the results presented in Tables~\ref{table:Numerical_Results_Table_1} and
\ref{table:Numerical_Results_Table_2}.

\begin{figure}[h!]
  \begin{subfigure}[b]{0.49\textwidth}
    \includegraphics[width=\textwidth]{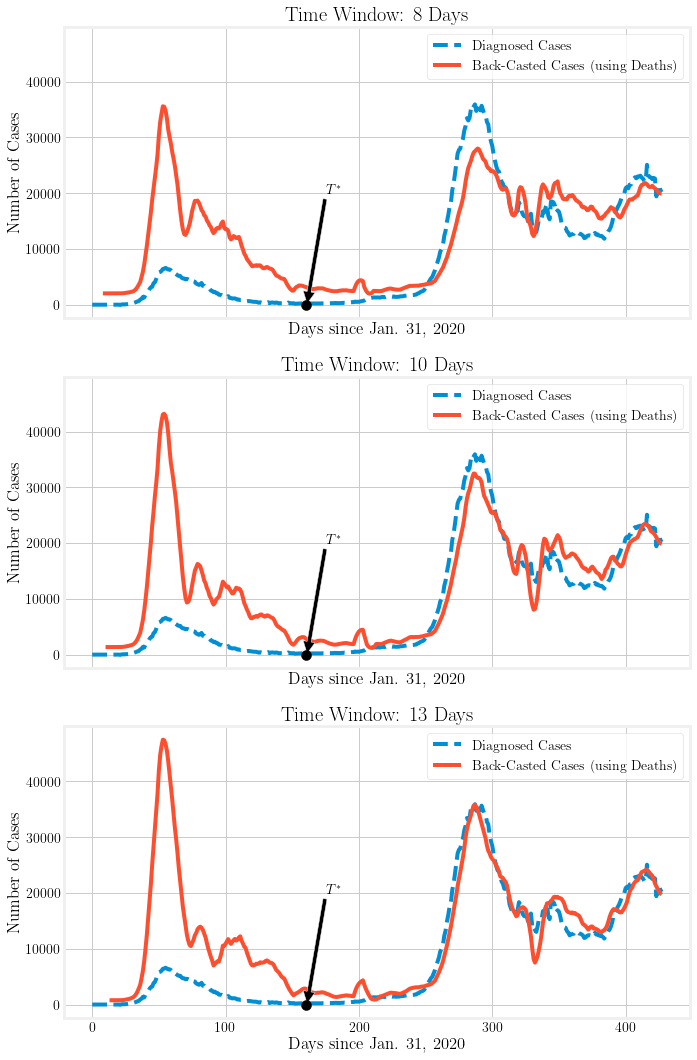}
    \caption{}
    \label{fig:time_window_comparison_italy}
  \end{subfigure}
  \hfill
  \begin{subfigure}[b]{0.49\textwidth}
    \includegraphics[width=\textwidth]{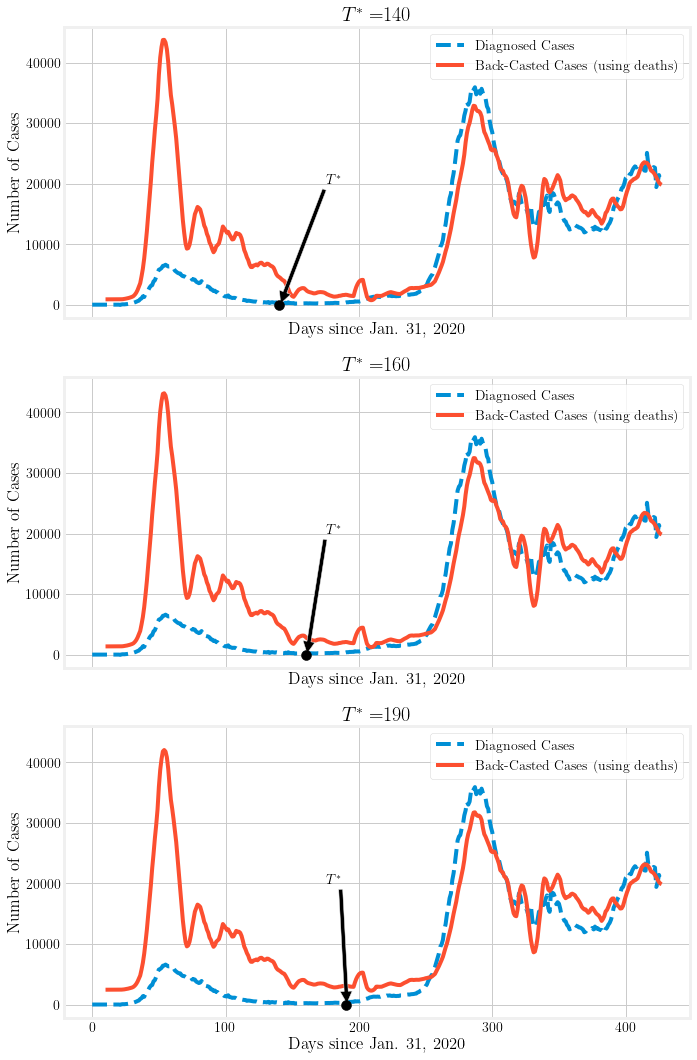}
    \caption{}
    \label{fig:t_star_comparison_italy}
  \end{subfigure}
  \caption{{Left panel:} The qualitative features of the GP regression are similar across different time windows (embedding dimensions). {Right panel:} Different values of $T^*$ for a fixed {embedding dimension} (time window) have minimal effect in the backcasted result. Note that after $T^*$ the observed values fall within the confidence interval around the predictions (which is not depicted).}
\end{figure}

We demonstrate the effect 
of changing the time window on the backcasting result {in Figure~\ref{fig:time_window_comparison_italy}},
{whereas} the effect of varying the $T^*$  {is shown in Figure~\ref{fig:t_star_comparison_italy}.}
In the latter it can be seen that variations
of $T^*$ do not change the qualitative
picture described above.
In the former, we can see that $k=13$
may lead to potential overfitting, while
already $k=10$ yields an adequate qualitative
representation of the number of cases.

\begin{figure}[h!]
    \centering
    \includegraphics[width=13cm]{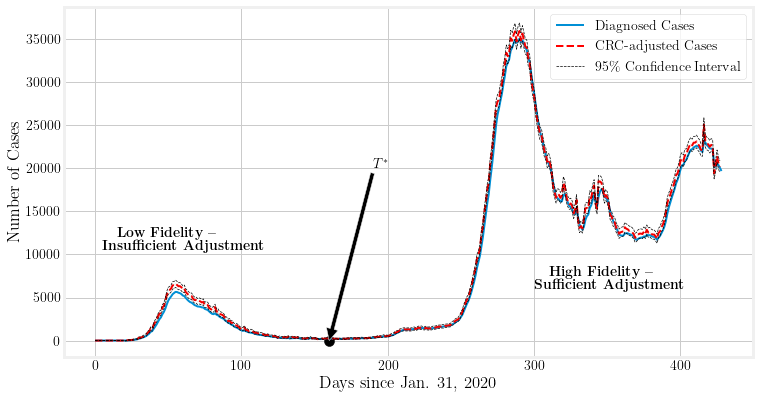}
    \caption{Demonstration of the capture-recapture (CRC) approach \cite{bohning2020} to estimate uncertainty in the diagnosed cases. The CRC adjustment (presented in \ref{sec:uncertainty_in_cases}) to the number of cases is insufficient when the estimated number of cases is not realistic, which can be seen in comparison to the backcasted estimate of Figure \ref{fig:Italy_BC_Daily}. However, they may be considered a reasonable estimate of the true number of missed cases later on ($\hat{c}_t$ for $t\geq T^*$).}
    \label{fig:crc_italy_demo}
\end{figure}

\begin{figure}[h!]
  \begin{subfigure}[b]{0.49\textwidth}
    \includegraphics[width=\textwidth]{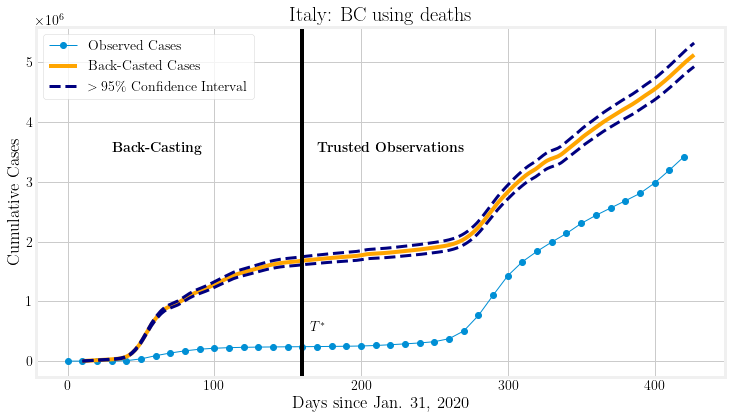}
    \caption{}
    \label{fig:crc_italy_backcastingA}
  \end{subfigure}
  \hfill
  \begin{subfigure}[b]{0.49\textwidth}
    \includegraphics[width=\textwidth]{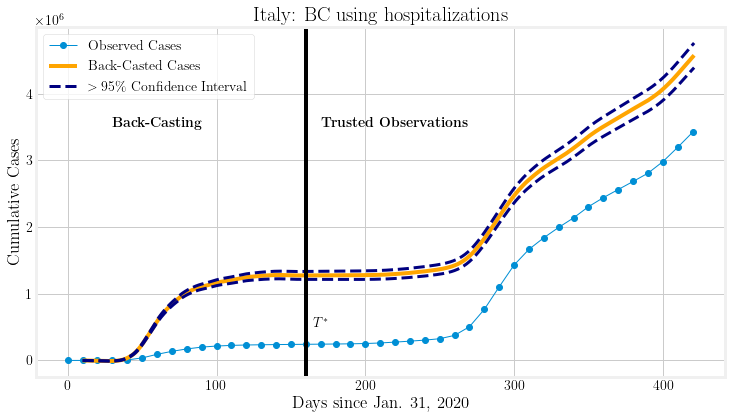}
    \caption{}
    \label{fig:crc_italy_backcastingB}
  \end{subfigure}
  \caption{Backcasting for Italy using the CRC case estimate and uncertainty after $T^*=160$.  Here $T^*=160, k=10$. 
  The left panel uses the data for deaths to perform the backcasting,
  while the right panel uses that for hospitalizations.
  While the results are not identical they are reasonably consistent.}
\end{figure}

 The uncertainty estimate associated with the GP implementation is not necessarily accurate, since the obtained uncertainty intervals can be tuned using the kernel parameters. To avoid additional computation, we circumvent this issue using the same $T^*=160$ and time window $k=10$, but apply the CRC method to adjust cases and their uncertainty after $T^*$ ({as described in} Section \ref{sec:uncertainty_in_cases}). {The associated backcasted results are presented in Figure~\ref{fig:crc_italy_demo}.} {Moreover, the cumulative number of cases based on the CRC-adjusted backcasting is shown in Figure~\ref{fig:crc_italy_backcastingA}.  Figure~\ref{fig:crc_italy_backcastingB}, instead, shows backcasting results} using the hospitalization time series in place of the fatalities time series as the predictor for cases. It is important to realize that the CRC adjustment of the cases can only be relied upon for $t \geq T^*$ and not where the estimated number of cases is unrealistic. We note similar behavior to Figure \ref{fig:Italy_BC_Cumulative} but with a much narrower uncertainty envelope due to the smaller uncertainty after the CRC adjustment.

\subsection{Comparative Results}
\label{main_results}

We extend our considerations to Spain, Sweden (where official policies related to COVID-19 were notably very different from those in other EU countries), Germany, United Kingdom, France
and South Korea, as well as Brazil, with the last two representing rather
extremely opposite-end examples of mitigation approaches towards the pandemic, as discussed below.
These countries (with the  exception of Brazil) fit the time-series 
profile described in Section \ref{sec:structure_and_t_star}: they have a first peak, 
followed by a period over the summer of 2020 where reported cases ebbed and 
then a subsequent second peak during the fall months of 2020. 

Among these countries, Sweden has the distinction of not ordering a nationwide lockdown during  the first wave  of the pandemic, as the other European countries considered in this work did \cite{flaxman2020}. Instead, among the mandated measures adopted were self isolation, social distancing, banning of public events
and partial school closures.
South Korea adopted a proactive approach and started developing testing 
capabilities almost two weeks before the first case in the country
\cite{oh2020}, established contact tracing protocols as early as mid-February of 2020, enforced social distancing from March 22 to April 15, demanded 
tests for all incoming travelers and quarantine for travelers from selected countries, and lastly, redistributed resources at hospitals and emphasized the use of personal protective
equipment by health care workers.

In sharp contrast, the lack of coordination at the federal level 
and delays in the implementation of mitigation measures in Brazil 
have been well documented \cite{buss2021, castro2021}. 
Testing capacity in the country was very 
limited, with test kits being  available 
for the first time in March 
2020, whereas the first phase of the epidemic in Brazil started in the 
middle of February 2020. Additionally, there were regions with 
extremely low ICU bed capacity, such as Amazonas, where the capacity was
11 beds for 100,000 people. As a result, Manaus, the capital city of 
Amazonas, was one of the regions that were affected particularly intensely.
For instance, fatalities per 100,000 people due to COVID-19 in this region exceeded 
by a factor of two the fatalities in the country overall. 

\begin{figure}[h]
    \centering
    \includegraphics[width=15cm]{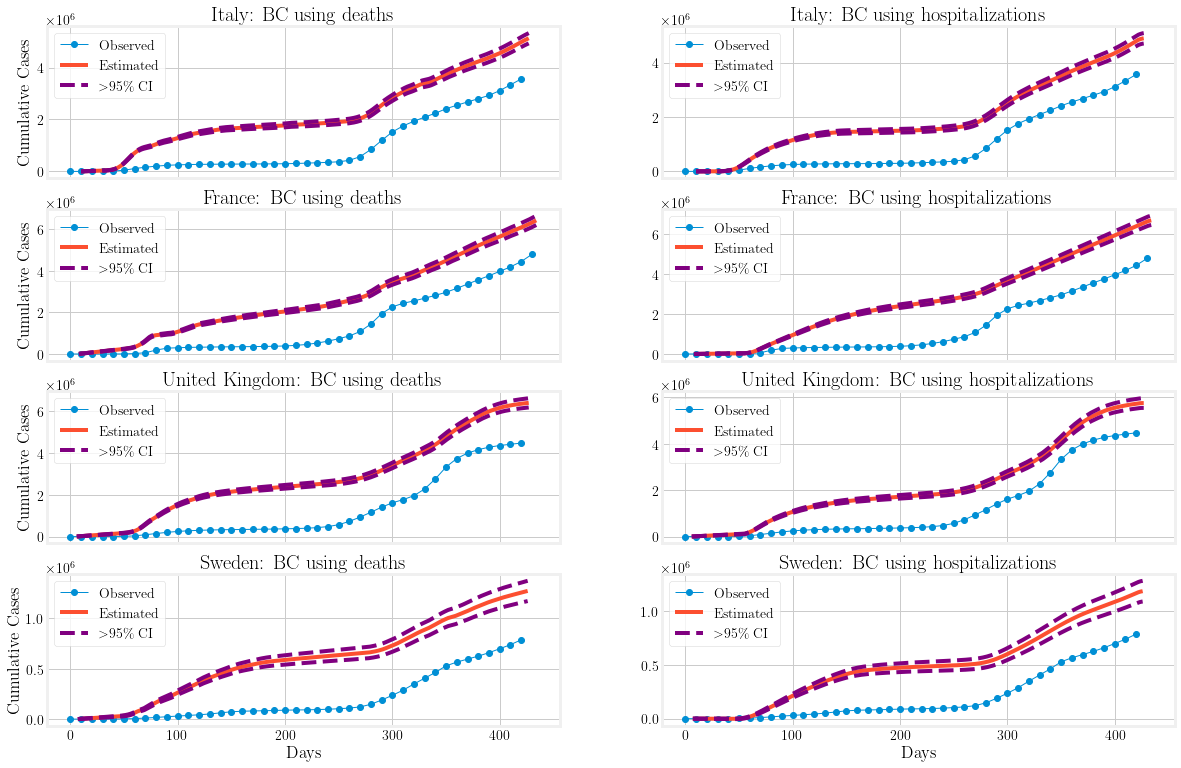}
    \caption{Backcasting {of the cumulative number of cases} using deaths (left) and hospitalizations (right) as predictors. For all of the countries we use $T^*=160$ and the corresponding embedding dimension estimated in Table \ref{tab:embedding_dimension_estimates}, 
    {as well a delay time of one day.}}
    \label{fig:numerical_results_1}
\end{figure}

All country time series (except that of Brazil, as previously mentioned)
satisfy the structure of Section \ref{sec:structure_and_t_star} (see, also, Figures \ref{fig:structure_of_cases} and \ref{fig:structure_of_deaths}) and this structure extends to the corresponding hospitalization data, when available. In the results {shown in Figure~\ref{fig:numerical_results_1}} we use both $D$ and $H$ as predictors, with the uncertainty estimated using the capture-recapture process \cite{bohning2020}. {The embedding dimension $k$ was again 
estimated using the False Nearest Neighbor algorithm,
with the results presented in Table~\ref{tab:embedding_dimension_estimates}.}
We find that the predicted, cumulative number of cases 
in the four countries illustrated in Figure~\ref{fig:numerical_results_1} was larger by a factor typically
between 1.3 and 2 when evaluated at the end time of our data set (April 4, 2021). However, the same factors vary greatly when computed at time $T^*$, at which point the estimate is purely through the backcasted result.

\begin{table}[h!]
    \centering
    \caption{Estimates of the embedding dimension $k$ of the death time series for each country.}
    \begin{tabular}{c|c|c}\hline \hline
        Country& Embedding dimension (Estimate, FNN)& Start Date (dd/mm/yy)\\
         \hline
         Brazil &5 &26/02/20\\
         France &8 &24/01/20\\
         Germany &9 &27/01/20\\
         Italy &10 &31/01/20\\
         South Korea &12 &21/01/20\\
         Spain &5 & 01/02/20\\
         Sweden &7 &01/02/20\\
         United Kingdom&6 &31/01/20 \\ \hline \hline
    \end{tabular}
    
    \label{tab:embedding_dimension_estimates}
\end{table}

\begin{table}[h!]
\centering
    
    \caption{Summary of the Fig. \ref{fig:numerical_results_1}
    backcasting predictions for the cumulative number of diagnosed cases at
    at time $T^*$ (July 9, 2020) and at the end of the considered period (April 4, 2021). Two different predictors were used: Death or
    hospitalization times series, both coupled with capture-recapture (CRC) uncertainty estimates.}
    
    \begin{adjustbox}{width=\textwidth}

    \begin{tabular}{c|c|cc|cc}
    \hline \hline
    \multicolumn{6}{c}{Backcasted cumulative number of cases ($>95$\% CI)} \\
    \hline
     Country  & Reported number & \multicolumn{2}{c|}{Death backcasting}  & \multicolumn{2}{c}{Hospitalization backcasting} \\
    & of cases $\times 10^{6}$ & Number $\times 10^{6}$& Backcasted to reported & Number $\times 10^{6}$ & Backcasted to reported \\
    \hline
    \multicolumn{6}{c}{09/07/2020 -- $T^*$}\\
    \hline
    Italy & 0.24 & 1.68 (1.61, 1.75) & 7.01 (6.73, 7.29) & 1.46 (1.40, 1.52) & 6.08 (5.83, 6.34) \\
    France & 0.20 & 1.89 (1.82, 1.96) & 9.27 (8.93, 9.61) & 2.17 (2.09, 2.25)& 10.66 (10.27, 11.04) \\
    Sweden & 0.08 & 0.61 (0.56, 0.66) & 7.97 (7.35, 8.58) & 0.48 (0.44, 0.53) & 6.33 (5.80, 6.87) \\
    United Kingdom & 0.28 & 2.20 (2.13, 2.28) & 7.73 (7.46, 8.01) & 1.57 (1.50, 1.64) & 5.51 (5.28, 5.74) \\
    \hline\hline
    \multicolumn{6}{c}{04/04/2021 -- Final Date}\\
    \hline
    Italy & 3.59 & 5.13 (5.93, 5.33) & 1.43 (1.37, 1.48) & 4.89 (4.70, 5.08) & 1.36 (1.31, 1.42) \\
    France & 4.70 & 6.54 (6.31, 6.77) & 1.39 (1.34, 1.44) & 6.84 (6.60, 7.08)& 1.45 (1.40, 1.51) \\
    Sweden & 0.80 & 1.35 (1.25, 1.46) & 1.69 (1.56, 1.82) & 1.22 (1.12, 1.32) & 1.53 (1.40, 1.65) \\
    United Kingdom & 4.36 & 6.40 (6.17, 6.62) & 1.47 (1.42, 1.52) & 5.75 (5.54, 5.96) & 1.32 (1.27, 1.37) \\
    \hline \hline
    \end{tabular}
    \end{adjustbox}
    \label{table:Numerical_Results_Table_1}
\end{table}

\begin{figure}[h]
    \centering
    \includegraphics[width=15cm]{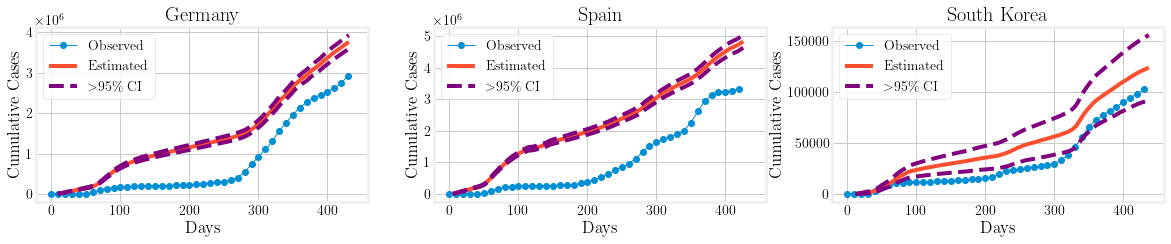}
    \caption{Backcasting results for countries where daily hospitalization records are not (or are partially) available. For all the countries we use $T^*=160$ and a window, embedding dimension, of $15$ days, and the same kernel for the Gaussian Process
    (Mat\'{e}rn kernel).}
    \label{fig:numerical_results_2}
\end{figure}

For Germany, Spain, South Korea, and Brazil, daily hospital occupancy is not readily available in our data set, and we only include the prediction based on the corresponding fatalities. {Figure~\ref{fig:numerical_results_2}
presents the backcasted cumulative number of cases (in analogy to the left panel of Figure~\ref{fig:numerical_results_1}), whereas Table~\ref{table:Numerical_Results_Table_2} presents the results of our calculations at $t=400$ and compares them to the reported number of cases (analogously with Table~\ref{table:Numerical_Results_Table_1}).}

\begin{table}[h!]
    \centering
    \textcolor{black}{
     \caption{Summary of the Fig. \ref{fig:numerical_results_2}
    backcasting predictions for the cumulative number of diagnosed cases at time $T^*$ (July 9, 2020) and at the end of the considered period (April 4, 2021). The death times series coupled with capture-recapture (CRC) uncertainty estimates was used as a predictor.}
    \begin{tabular}{c|c|cc}
     \hline \hline
    \multicolumn{4}{c}{Backcasted cumulative number of cases ($>95$\% CI)}\\
    \hline
      Country & Reported number & \multicolumn{2}{c}{Death backcasting} \\
      & of cases $\times 10^{6}$ & Number $\times 10^{6}$ & Backcasted to reported \\
    \hline
    \multicolumn{4}{c}{09/07/2020 -- $T^*$}\\
    \hline
    Germany & 0.19 & 1.03 (0.98, 1.08) &  5.30 (5.03, 5.58) \\
    South Korea & 0.01 & 0.03 (0.02, 0.4) & 2.64 (1.83, 3.46) \\
    Spain & 0.29 & 2.12 (2.03, 2.21)& 7.38 (7.06, 7.70)\\
    \hline \hline
    \multicolumn{4}{c}{04/04/2021 -- Final Date}\\
    \hline
    Germany & 2.84 & 3.79 (3.62,3.96) &  1.33 (1.27, 1.40) \\
    South Korea & 0.10 & 0.13 (0.09, 0.16) & 1.22 (0.91, 1.53) \\
    Spain & 3.28 & 5.17 (4.95, 5.38)& 1.57 (1.51, 1.64)\\
    \hline \hline
    \end{tabular}
    \label{table:Numerical_Results_Table_2}
    }
\end{table}

Two other studies that performed backcasting of cases are 
\cite{bohning2020} and \cite{flaxman2020}. According to
\cite{bohning2020}, which used a capture-recapture method, 
the ratio of total to {reported} cases for Italy, Germany, Spain and the UK, were
2.23, 2.30, 2.21 and 2.37, respectively. 
{Instead, according to the results shown in Tables~\ref{table:Numerical_Results_Table_1} and \ref{table:Numerical_Results_Table_2}, we find that the ratio of
backcasted cumulative number of cases to the reported number is
1.43 (Italy), 1.85 (Germany), 2.00 (Spain), and 1.47 (United Kingdom). These numbers are smaller than those reported in \cite{bohning2020}, but they are consistent with the overall observation that the number of reported cases during the first epidemic wave is significantly smaller than the actual number of infected individuals.} In \cite{flaxman2020},
backcasting was performed from observed deaths based on a Bayesian 
mechanistic model; they found that until May 4, 2020, the percentage of population infected in 
Italy, France, UK, Germany,  Sweden and Spain was
4.6 \%, 3.4\%, 5.1\%, 0.85\%, 3.7\% and 5.5\%,
respectively. These far exceed the reported infections in each of these countries. 

In comparison, our results in the same order are 2.8 \%, 2.8\%, 3.3\%, 1.2\%, 5.9\%, 4.5\%.
We note that these results are in the same ballpark as those of~\cite{flaxman2020},
although there are differences due to the detailed assumptions of each setting.

\begin{figure}[h]
  \begin{subfigure}[b]{0.49\textwidth}
    \includegraphics[width=\textwidth]{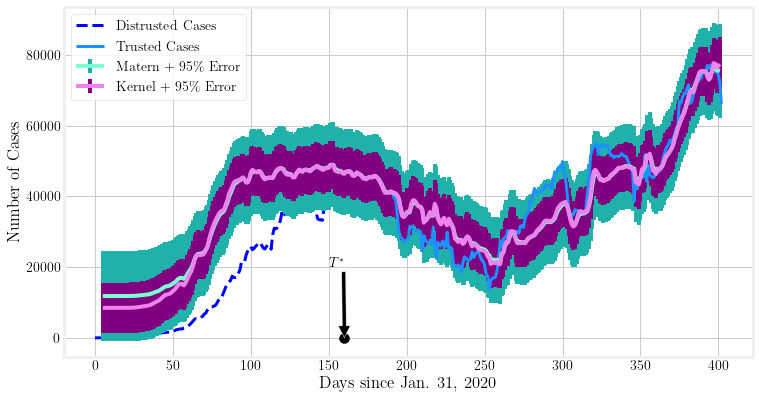}
    \caption{}
    \label{fig:Brazil_Daily_Wrong}
  \end{subfigure}
  \hfill
  \begin{subfigure}[b]{0.49\textwidth}
    \includegraphics[width=\textwidth]{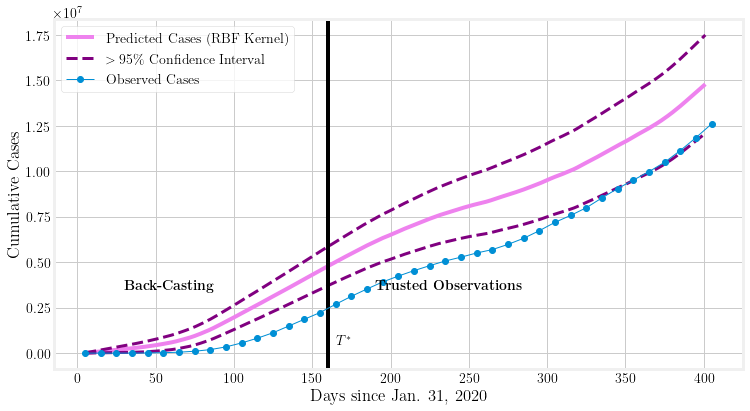}
    \caption{}
    \label{fig:Brazil_Cumulative}
  \end{subfigure}
  \caption{Backcasting results for Brazil with a time window of 5 days. The left panel includes results for two different choices of kernel (Exponential and RBF) for the GP regression.
  After time $T^*=160$, the predicted cases (pink) are close to the observed cases (blue) since the model has been trained at those points.
  The right panel only presents one of the results (RBF) for the sake of clarity.}
\end{figure}

We also include the backcasting projection for Brazil: {the number of daily cases is shown in Figure~\ref{fig:Brazil_Daily_Wrong}, whereas the cumulative number of cases is presented in Figure~\ref{fig:Brazil_Cumulative}}. There, the data do not satisfy the distinct wave structure of the previous examples. Irrespective of that, we found similar features as in the previously reported examples. In particular, we found that the diagnosed cases for $t\leq T^{\star}$ were under-reported although by a substantially smaller fraction than in other countries. Specifically, the proposed backcasting method predicts a higher and earlier peak in cases than observed, but it notably fails when trying to extrapolate outside of the range of the given data: early predictions, when cases and deaths due to Covid-19 are near zero, should yield a similar result. The left tail should therefore be considered inaccurate. 
We only include this example to demonstrate why the backcasting result may not be trustworthy if the structure of the data is not consistent with Section \ref{sec:structure_and_t_star}, even if it seems plausible.

\section{Discussion-Conclusion}
\label{sec:Discussion and Conclusion}

We developed a computational framework that uses as input the case and fatality incidence in the second wave of the COVID-19 epidemic in various regions to evaluate the magnitude of case incidence of the first wave, assuming the incidence  of fatalities is accurate. The  framework is based on algorithms that leverage the Takens-Whitney delay-embedding theorems \cite{sauer1991, takens1981, whitney1936}, use the method of False Nearest Neighbors \cite{abarbanel1993} to estimate the embedding dimension, and employ Gaussian Processes (GP) to perform nonlinear regression (as well as to quantify the resulting uncertainty). Uncertainty is also quantified using the capture-recapture approach of \cite{bohning2020}.
    
The method requires a minimal number of external parameters, besides those hyperparameters internal to the GP algorithm. These parameters include the embedding dimension $k$, the parameter $T^*$, and the time delay $\tau$, which for the 7-day averaged time series was taken to be one day. The parameter $k$ is the number of lagged observations required to construct the manifold $\mathcal{M}_{D}$ diffeomorphic to $\mathcal{M}$. The time parameter $T^*$ separates (somewhat artificially) the data into a segment past 
$T^*$, after which the data are considered to be accurate (typically a low incidence time during the summer). Assuming the accuracy of fatalities (or hospitalizations) the aim is to backcast the incidences during the first wave, i.e., during the early stages of the pandemic, based on the trends between cases and fatalities. The relatively small number of parameters and hyperparameters sets the current methodology apart from other ODE or PDE based approaches, for which a significant effort is required to resolve parameter identifiability issues \cite{cuevas2021, kevrekidis2021}. 
    
The framework was used with data from various European countries, which adopted disparate mitigation strategies, as well as with data from South Korea and Brazil, which also followed antipodal approaches to curtail the burden of the pandemic. Among the findings of this study is that in the six European countries studied (Italy, France, United Kingdom, Sweden, Germany and Spain) the first wave was, as widely suspected, considerably larger than reported.
{Specifically, at the end of the period considered, the ratio of predicted (backcasted) cumulative number of infections to those reported was found to be 1.43 (Italy), 1.39 (France), 1.47 (UK), 1.69 (Sweden), 1.33 (Germany), and 1.57 (Spain), i.e., the predicted number of cases was consistently greater by over 30\% than the reported number}. In South Korea, however, where the epidemic  was controlled, the discrepancy between predicted and observed cases is substantially smaller; we found the ratio of predicted to reported to be 1.20, a discrepancy of 20\%. This is a case example, as was explained in the text, where a significantly different (and far more proactive) mitigation strategy was brought to bear, which is apparently reflected in our results. 
    
There are numerous technical issues and further possibilities to consider along this line of efforts.
 
Backcasting can be implemented between other time series, e.g., between deaths and hospitalizations.
A relevant time series that has received considerable attention is that of ``excess deaths". This has been a traditional way to gauge the uncertainty around the death toll of exceptional events, such as the current pandemic \cite{weinberger2020} or extreme heat waves \cite{fouillet2008}. Excess deaths measure the deviation of the observed death toll from the expected death toll during a period of time, where the expected death toll is often a function (if not the direct average) of the death toll of the previous few years.
 
While this is an accurate measure of the overall impact of the pandemic, it takes into account increases or decreases in mortality due to other causes as well. Because of the disruptive nature of the pandemic, it is almost certain that the excess deaths seen are not all attributed to the virus itself: for example,
the diversion of most medical resources to the
care for COVID-19 patients, decreased their availability for the care
of patients with other chronic illnesses or with non-COVID-19 acute symptoms. There are also countries that have not seen a rise in excess deaths despite confirmed deaths due to the coronavirus \cite{karlinsky2021}. 
Australia and New Zealand recorded 
lower mortality than in recent years, which is attributed  to social
distancing measures that decreased mortality for reasons other than COVID-19, e.g., due to low incidence of seasonal flu. Uruguay and  Norway, also reported negative excess deaths. 

Trying to isolate true COVID-19 fatalities missed in the official numbers is a daunting (if not impossible) task, and that is acknowledged in the relevant literature as well \cite{beaney2020}.
Statistically, (and if we disregard the concerns of the above paragraph) we can perform regression between the observed and excess deaths to find what percent of deaths was missed. This is a process that must be done individually for each country, and its accuracy depends on the availability and quality of data sets, accumulated over several years.

An alternative method to quantify the magnitude of the true number of cases, which has been put in practice in various regions, is to monitor the viral load carried in sewer waste \cite{KostoglouWaste2022, scott2021}. It would be useful to examine the relevant correlations in subsequent waves vs. the measurements during the first wave. In a scenario where significant testing takes place (as, e.g., is the case in South Korea among our selected examples) or was subsequently developed, perhaps one can use the correlation between fatalities, symptomatic and asymptomatic cases, in order to infer the fractions of infections that stem from symptomatic vs. asymptomatic cases. This is an interesting aspect of such epidemiological models for which presently there is considerable uncertainty with very different asymptomatic fractions being reported in different studies \cite{li2020}.

\subsection*{Funding}
This work was funded by C3.ai Inc. and Microsoft Corporation. Additional support through the NSF (grant DMS-1815764 to Z.R.) is acknowledged. J.C.-M. acknowledges support from EU (FEDER program 2014- 2020) through both Consejería de Economía, Conocimiento, Empresas y Universidad de la Junta de Andalucía (under the projects P18-RT-3480 and US-1380977), and MICINN and AEI (under the projects PID2019-110430GB-C21 and PID2020-112620GB-I00).

\subsection*{Acknowledgements}
The views expressed are purely those of the authors and may not
in any circumstances be regarded as stating an official position of the European Commission.

\clearpage
\bibliographystyle{plain}
\bibliography{backcast}

\begin{thebibliography}{10}

\bibitem{abarbanel1993}
H.~D.~I. Abarbanel and M.~B. Kennel.
\newblock Local false nearest neighbors and dynamical dimensions from observed
  chaotic data.
\newblock {\em Physical Review E}, 47:3057--3068, 1993.

\bibitem{siettos2020}
C.~Anastassopoulou, L.~Russo, A.~Tsakris, and C.~Siettos.
\newblock Data-based analysis, modelling and forecasting of the
  {C}{O}{V}{I}{D}-19 outbreak.
\newblock {\em PLoS ONE}, 15(3):e0230405, 2020.

\bibitem{beaney2020}
T.~Beaney, J.~M. Clarke, V.~Jain, A.~K. Golestaneh, G.~Lyons, D.~Salman, and
  A.~Majeed.
\newblock Excess mortality: the gold standard in measuring the impact of
  {C}{O}{V}{I}{D}-19 worldwide?
\newblock {\em Journal of the Royal Society of Medicine}, 113(9):329--334,
  2020.

\bibitem{bohning2020}
D.~Bohning, I.~Rocchetti, A.~Maruotti, and H.~Holling.
\newblock Estimating the undetected infections in the {C}{O}{V}{I}{D}-19
  outbreak by harnessing capture–recapture methods.
\newblock {\em International Journal of Infectious Diseases}, 97:197--201,
  2020.

\bibitem{buss2021}
L.~F. Buss, C.~Prete~Jr, C.~M.~M. Abrahim, A.~Mendrone~Jr, T.~Salomon,
  C.~de~Almeida-Neto, R.~F.~O. Franca, M.~C. Belotti, M.~P. S.~S. Canrvalho,
  A.~G. Costa, M.~A.~E. Crispim, S.~C. Ferreira, N.~A. Frajil, S.~Gurzenda,
  C.~Whittaker, L.~T. Kamaura, P.~L. Takecian, P.~da~Silva~Peixoto, M.~K.
  Oikawa, A.~S. Sishiya, V.~Rocha, N.~A. Salles, A.~de~Souza~Santos, M.~A.
  da~Silva, B.~Custer, K.~V. Parag, M.~Barral-Netto, M.~U.~G. Kraemer, R.~H.~M.
  Pereira, O.~G. Pybus, M.~P. Busch, M.~C. Castro, C.~Dye, V.~H. Nascimento,
  N.~R. Faria, and E.~Sabino.
\newblock Three-quarters attack rate of {S}{A}{R}{S}-{C}o{V}-2 in the
  {B}razilian {A}mazon during a largely unmitigated epidemic.
\newblock {\em Science}, 371:288--6292, 2021.

\bibitem{cao1997afn}
Liangyue Cao.
\newblock Practical method for determining the minimum embedding dimension of a
  scalar time series.
\newblock {\em Physica D: Nonlinear Phenomena}, 110(1-2):43--50, 1997.

\bibitem{castro2021}
M.~C. Castro, S.~Kim, L.~Berberia, A.~F. Ribeiro, S.~Gurmenda, K.~B. Ribeiro,
  E.~Abbott, J.~Blossom, B.~Rache, and B.~H. Singer.
\newblock Spatiotemporal pattern of {C}{O}{V}{I}{D}-19 spread in {B}razil.
\newblock {\em Science}, 372:821--826, 2021.

\bibitem{ciminelli2020}
G.~Ciminelli and S.~Garcia-Mandico.
\newblock {C}{O}{V}{I}{D}-19 in {I}taly: An analysis of death registry data.
\newblock {\em Journal of Public Health}, 42(4):723--730, 2020.

\bibitem{cuevas2021}
J.~Cuevas-Maraver, P.G. Kevrekidis, Q.Y. Chen, G.A. Kevrekidis, Víctor
  Villalobos-Daniel, Z.~Rapti, and Y.~Drossinos.
\newblock Lockdown measures and their impact on single- and two-age-structured
  epidemic model for the {C}{O}{V}{I}{D}-19 outbreak in {M}exico.
\newblock {\em Mathematical Biosciences}, 336:108590, 2021.

\bibitem{deyle2011}
E.~R. Deyle and G.~Sugihara.
\newblock Generalized theorems for nonlinear state space reconstruction.
\newblock {\em PLoS ONE}, 6(3):e18295, 2011.

\bibitem{dong2020interactive}
Ensheng Dong, Hongru Du, and Lauren Gardner.
\newblock An interactive web-based dashboard to track {C}{O}{V}{I}{D}-19 in
  real time.
\newblock {\em The Lancet infectious diseases}, 20(5):533--534, 2020.

\bibitem{flaxman2020}
S.~Flaxman, S.~Mishra, A.~Gandy, H.~J.~T. Unwin, T.~A. Mellan, H.~Coupland,
  C.~Whittaker, H.~Zhu, T.~Berah, J.~W. Eaton., M.~Monod, Imperial College
  COVID-19~Response Team, A.~C. Ghani, C.~A. Donnelly, S.~Riley, M.A.C.
  Vollmer, N.M. Ferguson, L.~Okell, and S~Bhatt.
\newblock Estimating the effects of non-pharmaceutical interventions on
  {C}{O}{V}{I}{D}-19 in {E}urope.
\newblock {\em Nature}, 584:257--261, 2020.

\bibitem{cdc}
Center for Disease~Control and Prevention.
\newblock Commercial laboratory seroprevalence surveys.
\newblock
  https://www.cdc.gov/coronavirus/2019-ncov/cases-updates/commercial-lab-surveys.html.

\bibitem{fouillet2008}
A.~Fouillet, G.~Rey, V.~Wagner, K.~Laaidi, P.~Empereur-Bissonnet, A.~Le~Tertre,
  P.~Frayssinet, P.~Bessemoulin, F.~Laurent, P.~De~Crouy-Chanel, E.~Jougla, and
  D.~Henmon.
\newblock Has the impact of heat waves on mortality changed in {F}rance since
  the {E}uropean heat wave of summer 2003? a study of the 2006 heat wave.
\newblock {\em International Journal of Epidemiology}, 37:309--317, 2008.

\bibitem{fraser1986}
A.~M. Fraser and H.L. Swinney.
\newblock Independent coordinates for strange attractors from mutual
  information.
\newblock {\em Physical Review A}, 33:1134--1140, 1986.

\bibitem{gilpin2020deep}
William Gilpin.
\newblock Deep reconstruction of strange attractors from time series.
\newblock {\em arXiv preprint arXiv:2002.05909}, 2020.

\bibitem{GNANVI2021258}
Janyce~Eunice Gnanvi, Kolawolé~Valère Salako, Gaëtan~Brezesky Kotanmi, and
  Romain {Glèlè Kakaï}.
\newblock On the reliability of predictions on {C}{O}{V}{I}{D}-19 dynamics: {A}
  systematic and critical review of modelling techniques.
\newblock {\em Infectious Disease Modelling}, 6:258--272, 2021.

\bibitem{hasell2020}
J.~Hasell, E.~Mathieu, D.~Beltekian, B.~Macdonald, C.~Giattino,
  E.~Ortiz-Ospina, M.~Roser, and H.~Ritchie.
\newblock A cross-country database of {C}{O}{V}{I}{D}-19 testing.
\newblock {\em Scientific Data}, 7:345, 2020.

\bibitem{holmdahl2020}
I.~Holmdahl and C.~Buckee.
\newblock Wrong but useful- what {C}{O}{V}{I}{D}-19 epidemiological models can
  and cannot tell us.
\newblock {\em The New England Journal of Medicine}, 383(4):303--305, 2020.

\bibitem{ing2020}
A.~J. Ing, C.~Cocks, and J.~P. Green.
\newblock {C}{O}{V}{I}{D}-19: in the footsteps of {E}rnest {S}hackleton.
\newblock {\em Thorax}, 75:693--694, 2020.

\bibitem{doi:10.1177/0272989X21990391}
Lyndon~P. James, Joshua~A. Salomon, Caroline~O. Buckee, and Nicolas~A. Menzies.
\newblock The use and misuse of mathematical modeling for infectious disease
  policymaking: {L}essons for the {C}{O}{V}{I}{D}-19 pandemic.
\newblock {\em Medical Decision Making}, 41(4):379--385, 2021.
\newblock PMID: 33535889.

\bibitem{james2021}
N.~James, M.~Menzies, and P.~Radchenko.
\newblock {C}{O}{V}{I}{D}-19 second wave mortality in {E}urope and the {U}nited
  {S}tates.
\newblock {\em Chaos}, 31:031105, 2021.

\bibitem{karlinsky2021}
A.~Karlinsky and D.~Kobak.
\newblock Tracking excess mortality across countries during the
  {C}{O}{V}{I}{D}-19 pandemic with the {W}orld {M}ortality {D}ataset.
\newblock {\em eLife}, 10:e69336, 2021.

\bibitem{kevrekidis2021}
P.~G. Kevrekidis, J.~Cuevas-Maraver, Y.~Drossinos, Z.~Rapti, and G.~A.
  Kevrekidis.
\newblock Reaction-diffusion spatial modeling of {C}{O}{V}{I}{D}-19: {G}reece
  and {A}ndalusia as case examples.
\newblock {\em Physical Review E}, 104:024412, 2021.

\bibitem{kolokolnikov2021}
T.~Kolokolnikov and D.~Iron.
\newblock Law of mass action and saturation in sir model with application to
  coronavirus modelling.
\newblock {\em Infectious Disease Modelling}, 6:91--97, 2021.

\bibitem{krakovska2015use}
Anna Krakovsk{\'a}, Krist{\'\i}na Mezeiov{\'a}, and Hana
  Bud{\'a}{\v{c}}ov{\'a}.
\newblock Use of false nearest neighbours for selecting variables and embedding
  parameters for state space reconstruction.
\newblock {\em Journal of Complex Systems}, 2015, 2015.

\bibitem{lan2021}
F.-Y. Lan, R.~Filler, S.~Mathew, E.~Iliaki, R.~Osgood, l.~A. Bruno-Murtha, and
  S.~N. Kales.
\newblock Evolving virulence? decreasing {C}{O}{V}{I}{D}-19 complications among
  {M}assachusetts healthcare workers: a cohort study.
\newblock {\em Pathogens and Global Health}, 115(1):4--6, 2021.

\bibitem{lee2019linking}
Seungjoon Lee, Felix Dietrich, George~E Karniadakis, and Ioannis~G Kevrekidis.
\newblock Linking {G}aussian process regression with data-driven manifold
  embeddings for nonlinear data fusion.
\newblock {\em Interface {F}ocus}, 9(3):20180083, 2019.

\bibitem{li2020}
R.~Li, S.~Pei, B.~Chen, Y.~Song, T.~Zhang, W.~Yang, and J.~Shaman.
\newblock Substantial undocumented infections facilitates the rapid
  dissemination of novel coronavirus ({S}{A}{R}{S}-{C}o{V}-2).
\newblock {\em Science}, 368:489--493, 2020.

\bibitem{liu2020}
Y.~Liu, A.~A. Gayle, A.~Wilder-Smith, and Rocklov J.
\newblock The reproduction number of {C}{O}{V}{I}{D}-19 is higher compared to
  {S}{A}{R}{S} coronavirus.
\newblock {\em Journal of Travel Medicine}, 2020.

\bibitem{mammeri2020}
Y.~Mammeri.
\newblock A reaction-diffusion system to better comprehend the unlockdown:
  Application of {S}{E}{I}{R}-type model with diffusion to the spatial spread
  of {C}{O}{V}{I}{D}-19 in {F}rance.
\newblock {\em Computational and Mathematical Biophysics}, 8:102--113, 2020.

\bibitem{Mannattil_2016_nolitsa}
Manu Mannattil, Himanshu Gupta, and Sagar Chakraborty.
\newblock Revisiting evidence of chaos in {X}-{R}ay light curves: {T}he case of
  {GRS} 1915$\mathplus$105.
\newblock {\em The Astrophysical Journal}, 833(2):208, dec 2016.

\bibitem{mannattil2017applicability_nolitsa}
Manu Mannattil, Ambrish Pandey, Mahendra~K Verma, and Sagar Chakraborty.
\newblock On the applicability of low-dimensional models for convective flow
  reversals at extreme {P}randtl numbers.
\newblock {\em The European Physical Journal B}, 90(12):1--9, 2017.

\bibitem{mathieu}
E.~Mathieu, H.~Richie, E.~Ortiz-Opsina, M.~Roser, J.~Hasell, C.~Appel,
  C.~Giattino, and L.~Rod{\'e}s-Guirao.
\newblock A global database of {C}{O}{V}{I}{D}-19 vaccinations.
\newblock {\em Nature Human Behaviour}, 2021.

\bibitem{noakes1991}
L.~Noakes.
\newblock The {T}akens embedding theorem.
\newblock {\em International Journal of Bifurcation and Chaos}, 1(4):867--872,
  1991.

\bibitem{oh2020}
J.~Oh, J.-K. Lee, D.~Schwarz, H.~L. Ratcliffe, J.~F. Markuns, and L.~R.
  Hirschhorn.
\newblock National response to {C}{O}{V}{I}{D}-19 in the {R}epublic of {K}orea
  and lessons learned for other countries.
\newblock {\em Health Systems and Reform}, 6:e1753464, 2020.

\bibitem{omori2020}
R.~Omori, K.~Mizumoto, and G.~Chowell.
\newblock Changes in testing rates could mask the novel coronavirus disease
  ({C}{O}{V}{I}{D}-19) growth rate.
\newblock {\em International Journal of Infectious Diseases}, 94:116--118,
  2020.

\bibitem{paixao2021}
B.~Paixao, L.~Baroni, M.~Pedroso, R.~Salles, L.~Escobar, R.~de~Sousa, C.and de
  Freitas~Saldanha, J.~Soares, R.~Coutinho, F.~Porto, and E.~Ogasawara.
\newblock Substantial underestimation of {S}{A}{R}{S}-{C}o{V}-2 infection in
  the {U}nited {S}tates.
\newblock {\em New Generation Computing}, 2021.

\bibitem{scikit-learn}
F.~Pedregosa, G.~Varoquaux, A.~Gramfort, V.~Michel, B.~Thirion, O.~Grisel,
  M.~Blondel, P.~Prettenhofer, R.~Weiss, V.~Dubourg, J.~Vanderplas, A.~Passos,
  D.~Cournapeau, M.~Brucher, M.~Perrot, and E.~Duchesnay.
\newblock Scikit-learn: Machine learning in {P}ython.
\newblock {\em Journal of Machine Learning Research}, 12:2825--2830, 2011.

\bibitem{bhatta2020}
M.~Peirlinck, K.~Linka, F.~Sahli~Costabal, J.~Bhattacharya, E.~Bendavid,
  J.~A.~P. Ioannidis, and E.~Kuhl.
\newblock Visualizing the invisible: The effect of asymptomatic transmission on
  the outbreak dynamics of {C}{O}{V}{I}{D}-19.
\newblock {\em Computer Methods in Applied Mechanics and Engineering}, 2020.

\bibitem{KostoglouWaste2022}
M.~Petala, M.~Kostoglou, Th. Karapantsios, C.I. Dovas, Th. Lytras,
  D.~Praskevis, E.~Roilides, A.~Koutsolioutsou-Benakih, G.~Panagiotakopoulos,
  V.~Sypsa, S.~Metallidis, A.~Papa, E.~Stylianidis, A.~Papadopoulos,
  S.~Tsiodras, and N.~Papaioannou.
\newblock Relating {S}{A}{R}{S}-{C}o{V}-2 shedding rate in wastewater to daily
  positive tests data: {A} consistent model based approach.
\newblock {\em Science of the Total Environment}, 807:150838, 2022.

\bibitem{phippsl2020}
S.~J. Phipps, R.~Q. Grafton, and T.~Kompas.
\newblock Robust estimates of the true (population) infection rate for
  {C}{O}{V}{I}{D}-19: a backcasting approach.
\newblock {\em Royal Society Open Science}, 7:200909, 2020.

\bibitem{rasmussen2006cki}
C.~E. Rasmussen and C.~K.~I. Williams.
\newblock Gaussian {P}rocesses for {M}achine {L}earning.
\newblock In T.~Dietterich, editor, {\em {A}daptive {C}omputation and {M}achine
  {L}earning}. {M}{I}{T} {P}ress, 55 Hayward Street, Cambridge, MA 02142, 2006.

\bibitem{russell2020}
T.~W Russell, N.~Golding, J.~Hellewell, S.~Abbott, L.~Wright, C.A.B. Pearson,
  K.~van Zandvoort, C.I. Jarvis, H.~Gibbs, Y.~Liu, R.~M. Eggo, W.~J. Edmunds,
  A.J. Kucharski, and {C}{M}{M}{I}{D} {C}{O}{V}{I}{D}-19 working group.
\newblock Reconstructing the early global dynamics of under-ascertained
  {C}{O}{V}{I}{D}-19 cases and infections.
\newblock {\em BMC Medicine}, 18:322, 2020.

\bibitem{sah2021}
Pratha Sah, Meagan~C. Fitzpatrick, Charlotte~F. Zimmer, Elaheh Abdollahi,
  Lyndon Juden-Kelly, Seyed~M. Moghadas, Burton~H. Singer, and Alison~P.
  Galvani.
\newblock Asymptomatic {S}{A}{R}{S}-{C}o{V} infection: A sustematic review and
  meta-analysis.
\newblock {\em Proceedings of the {N}ational {A}cademy of {S}ciences
  ({U}.{S}.{A})}, 118(34):e2109229118, 2021.

\bibitem{sauer1991}
T.~Sauer, J.~A. Yorke, and M.~Casdagli.
\newblock Embedology.
\newblock {\em Journal of Statistical Physics}, 65:579--616, 1991.

\bibitem{scott2021}
L.~C. Scott, A.~Aubee, L.~Babahaji, K.~Vigil, S.~Tims, and T.~G. Aw.
\newblock Targeted wastewater surveillance of {S}{A}{R}{S}-{C}o{V}-2 on a
  university campus for {C}{O}{V}{I}{D}-19 outbreak detection and mitigation.
\newblock {\em Environmental Research}, 200:111374, 2021.

\bibitem{stralin2021}
K.~Stralin, E.~Wahlstrom, S.~Walther, A.~M. Bennet-Bark, M.~Heurgren,
  T.~Linden, J.~Holm, and H.~Hanberger.
\newblock Mortality trends among hospitalised {C}{O}{V}{I}{D}-19 patients in
  {S}weden: A nationwide observational cohort study.
\newblock {\em The Lance Regional Health-Europe}, 4:100054, 2021.

\bibitem{takens1981}
F.~Takens.
\newblock Detecting strange attractors in turbulence.
\newblock {\em Lecture Notes in Mathematics}, 898:366--381, 1981.

\bibitem{viceconte2020}
G.~Viceconte and N.~Petrosillo.
\newblock {C}{O}{V}{I}{D}-19 {R}{0}: {M}agic number or conundrum?
\newblock {\em Infectious Disease Reports}, 12:8516, 2020.

\bibitem{viguerie2021}
A.~Viguerie, G.~Lorenzo, F.~Auricchio, D.~Baroli, T.~J.~R. Hughes, A.~Patton,
  A.~Reali, T.~E. Yankeelov, and A.~Veneziani.
\newblock Simulating the spread of {C}{O}{V}{I}{D}-19 via a spatially-resolved
  susceptible–exposed–infected–recovered–deceased ({S}{E}{I}{R}{D})
  model with heterogeneous diffusion.
\newblock {\em Applied Mathematics Letters}, 111:106617, 2021.

\bibitem{weinberger2020}
D.~M. Weinberger, J.~Chen, T.~Cohen, F.~W. Crawford, F.~Mostashari, D.~Olson,
  V.E. Pitzer, N.~G. Reich, M.~Russi, L.~Simonsen, A.~Watkins, and C.~Viboud.
\newblock Estimation of excess deaths associated with the {C}{O}{V}{I}{D}-19
  pandemic in the {U}nited {S}tates, {M}arch to {M}ay 2020.
\newblock {\em JAMA Internal Medicine}, 180(10):1336--1344, 2020.

\bibitem{whitney1936}
H.~Whitney.
\newblock Differentiable manifolds.
\newblock {\em Annals of Mathematics}, 37:645--680, 1936.

\bibitem{wu2020}
S.~L. Wu, A.~N. Mertens, Y.~S. Crider, A.~Nguyen, N.~N. Pokpongkiat,
  S.~Djajadi, A.~Seth, M.~S. Hsiang, A.~Colford~Jr., J. M.and~Reingold, B.~F.
  Arnold, A.~Hubbard, and J.~Benjamin-Chung.
\newblock Substantial underestimation of {S}{A}{R}{S}-{C}o{V}-2 infection in
  the {U}nited {S}tates.
\newblock {\em Nature Communications}, 11:4507, 2020.

\bibitem{zhao2020preliminary}
Shi Zhao, Qianyin Lin, Jinjun Ran, Salihu~S Musa, Guangpu Yang, Weiming Wang,
  Yijun Lou, Daozhou Gao, Lin Yang, Daihai He, et~al.
\newblock Preliminary estimation of the basic reproduction number of novel
  coronavirus (2019-n{C}o{V}) in {C}hina, from 2019 to 2020: A data-driven
  analysis in the early phase of the outbreak.
\newblock {\em International journal of infectious diseases}, 92:214--217,
  2020.

\end{thebibliography}

\clearpage
\appendix

\section{Data}
\label{sec:Sources}
\subsection{Source and Quality}
The  data  used  in  the Numerical Results, Section 3 in the main text,
come  from a  single  source  for  consistency. We  use  the  publicly available data set by Our-World-In-Data (OWID)\footnote{https://github.com/owid/covid-19-data} that  aggregates  information  from  other  trustworthy  sources (for confirmed cases and deaths the JHU database \cite{dong2020interactive} is used, while hospitalizations are sourced from several government websites). This means that availability and quality of some data categories are different across countries (for  example, Germany does not readily publish daily hospitalization data and Spain has days on which the reported cases are negative, functioning as adjustments to previous reports). In turn, comparisons between statistics across countries is not always possible, and even if it is, may not be trustworthy. 

Due to these and related issues, our estimates are dependent on data quality and trustworthiness. To avoid large disparities in reporting strategies and 
healthcare we have focused on a few selected countries. Furthermore, we address
data uncertainty in the following section.

The latest version of the data set used was downloaded on April 4, 2021. We consider data after the Spring of 2021 to correspond to a very different dynamical system, in which a large part of the population (of countries for which results are presented) almost discontinuously obtained immunity due to vaccine roll-out. Thus, we do not consider later data in this work.

Our model is centered around learning the relationship between diagnosed cases ($c_t$) and reported fatalities ($d_t$). Assuming that there is enough trustworthy information (past the selected $T^*$), we can extrapolate backwards to points in time where we do not believe the available information. However, even though COVID-19 monitoring has been improving continuously, it would be na\"{i}ve to entirely trust even the most recent information. We therefore attempt to address uncertainties that may surround the two types of time series we work with.

\subsection{Addressing Data Uncertainty}
Our model is centered around learning the relationship between diagnosed cases ($c_t$) and reported fatalities ($d_t$). Assuming that there is enough trustworthy information (past the selected $T^*$), we can extrapolate backwards to points in time where we do not believe the available information. However, even though COVID-19 monitoring has been improving continuously, it would be na\"{i}ve to entirely trust even the most recent information. We therefore attempt to address uncertainties that may surround the two types of time series we work with.

\subsubsection{Reliability of Time Series} 
Of all the pandemic markers available to us, fatalities and hospitalizations are possibly the most reliable data sets, as we have also discussed at some length in earlier work~\cite{cuevas2021}. Since the method developed in Section 2.1 (main text)
is valid for arbitrary time series, as long as these two are coupled in the dynamical system, we can replace the role of the death time series ($D$) by a hospitalization time series ($H$), under the reasonable assumption that the number of hospitalized patients should also be an observable quantity of the pandemic. 

In principle, one can use either time series (and any other similar observable, such as ICU admissions) to backcast the numbers of cases. This can be used as a tool to increase the certainty of a prediction by repeating the computational experiment using different time series. However, hospitalization data are less readily available world-wide, and for countries with low hospital capacity the time series may not be as predictive, since there is no way to quantify the severity of an outbreak solely on these records after capacity is reached; after that point, capacity saturation yields misleading results regarding the number of cases.

{It is important to note that our estimates of the true case incidence include a mixture of symptomatic \textit{and} asymptomatic cases. This is because early on tests were scarce and not available to the broader public, while in latter stages asymptomatic testing became broadly available, so many cases that were asymptomatic or mild would be included in the corresponding measurements. Early case numbers \textit{do not} in general contain that information, differentiating in this way this portion of the time series.}

\subsection{Hyperparameters}
The hyperparameters needed to perform the relevant computations, i.e., parameters that control the learning process but do not necessarily have an interpretable physical meaning, can be split in two groups: 

\begin{enumerate}
    \item Those associated with the reconstruction of the diffeomorphic manifold (namely the embedding dimension $k$ and the delay time $\tau$).
    \item Those given to the GP regressor (i.e, specifying the kernel).
\end{enumerate}

In this section we explain how these hyperparameters are set and their relevance to physical quantities and characteristics of the observed data, where relevant. 

\textbf{(1) Embedding Parameters}

$\tau$ is the \textit{delay time} between subsequent observations of $d_t$ needed when performing the $\mathcal{M}_D$ manifold reconstruction
(Appendix C). 
It is meant to capture a natural variance that may occur in the dynamical system over a shorter time scale, and can be estimated using empirical methods (e.g., based on mutual information, as in \cite{fraser1986}). However, because of the weekly fluctuation in our data (both in recorded cases and fatalities, which decrease substantially during the weekends), it is \textit{a priori} natural to smoothen the data using a 7-day rolling average. This has the effect of making $\tau=1$ day for all of our data sets which become mostly smooth, meaning that we can use sequential  observations of $d_t$ when constructing $\mathcal{M}_{D}$. Note that using the smoothed data is not a prerequisite for the proposed method.

The \textit{embedding dimension} $k$ is the number of sequential observations of $d_t$ needed to reconstruct $\mathcal{M}_D$, i.e., the dimension of $\mathcal{M}_D$. It can be estimated using empirical algorithms such as False Nearest Neighbors (FNN) \cite{abarbanel1993} and Average False Neighbors (AFN) \cite{cao1997afn}. This parameter need not be optimal: The theory presented in Section 2.1 (main text)
guarantees that an embedding dimension of at most $2n+1$ is sufficient to embed in Euclidean space, where $n$ is the dimension of the original manifold. If $k^*$, {the minimal embedding dimension,} is sufficient, reconstructions using $k>k^*$  points will still be diffeomorphic to $\mathcal{M}$ (in the absence of noise). Still, it is important, that $n$ is small enough so that the available data are adequate for the reconstruction: e.g., for $n=5$ we can take $k=11$ and a point $\vb{d}_t=\{d_{t-10},...,d_{t}\}\in\mathcal{M}_d\subset\mathbb{R}^{11}$. Then we will require that $\hat{f}_\text{GP}(\vb{d}_t)=c_t$ to be the cases at time $t$ (where $\hat{f}_\text{GP}$ is the transformation based on the GP estimator, see, also, Section 3.1). 
Thus, having a large $n$ corresponds to needing a longer time series to reconstruct the corresponding manifold. Furthermore, using a larger $k$ than necessary may lead to overfitting (\cite{krakovska2015use}, \cite{gilpin2020deep}). 
We may refer to the embedding dimension as the ``\textit{time-window}" which defines the dimension of a point in $\mathcal{M}_{D}$.

\textbf{(2) Gaussian Process Parameters}

Herein, we describe broadly a mathematical framework that
justifies the use of regression to learn the relationship between observed cases of and deaths due to COVID-19, in contrast to other methods. We use Gaussian Process regression 
to obtain the transformation $\hat{f}_\text{GP}$. The choice of GP \textit{for} regression is made because of the integrated uncertainty estimates the method offers.

The \textit{GP Kernel} is the family of functions used to fit the data, and thereby obtain $\hat{f}_\text{GP}$. We use the Mat\'{e}rn
(SM, B.3) 
kernel which is a generalization of the Radial Basis Function kernel (RBF) (SM, B.2).
This choice is made empirically based on the characteristics of the data. For COVID-19 series we observe relatively smooth trajectories with no periodic features and some noise. (This would not be the case if the data were not smoothed \textit{a priori}, as weekly periodicity in the un-smoothed data would warrant the addition of a periodic kernel). Though we only include results of another kernel (RBF) in Section~3.1,
we consistently observe that the choice of kernel need not be unique, and the backcasting projection is mainly dependent on the relationship between the observed data and not on the choice of kernel (assuming the latter is reasonable). For the equations and specific kernel parameters see Appendix A.

While we may use the terms ``parameter" and ``hyperparameter" interchangeably in this text, we note that their choice \textit{does not} raise identifiability issues, since it is indeed possible to recover the same accuracy of a resulting estimator with different choices of hyperparameters, and parameter uniqueness is not required by the nature of our problem (the pandemic). This is in contrast to closed form ODE/PDE models in which the values of corresponding parameters are fundamentally significant in interpreting results of such a model \cite{kevrekidis2021, mammeri2020, viguerie2021}.

\section{Gaussian Processes}
\label{sec:appB}
Gaussian process regression is a non-parametric and supervised learning methodology
that can be used to make predictions and simultaneously quantify uncertainty. 
Prior knowledge about the system is encoded in the probability kernels used and 
regression is performed by specifying distributions over an infinite number of 
possible functions passing through the observation points. 

A Gaussian process \cite{rasmussen2006cki} is specified by its mean ($m$) and covariance ($k$) functions. If $c = f(\vb{d})$ is the real process then:
\begin{align*}
    m(\vb{d_t}) &= \mathbb{E}\qty[f(d_t)]\\
    k(\vb{d_t},\vb{d_{t'}})&=\mathbb{E}[f(\vb{d_t})-m(\vb{d_t})]\mathbb{E}[f(\vb{d_{t'}})-m(\vb{d_{t'}})]
\end{align*}
and
\begin{align}
    f(\vb{d}_t)\sim\mathcal{GP}(m(\vb{d_t}), k(\vb{b_t},\vb{b_{t'}}))
\end{align}

In the \texttt{sklearn} implementation of the method the covariance matrix specified by the kernel function $k$ has a zero diagonal value, denoted by the parameter $\alpha$. There are numerical advantages to having a nonzero such value which makes the matrix positive definite. In the results of Section 3
we use the default $\alpha=1\text{e}-10$.

Overall we make use of two kernels: the Radial Basis Function Kernel (RBF) given by
\begin{align}
    k_\text{RBF}(x_i,x_j)=\exp(-\frac{\norm{x_i-x_j}_2^2}{2l^2})
    \label{eqn:RBF}
\end{align}

and the Mat\'{e}rn Kernel.
\begin{equation}
    k_M(x_i,x_j)=\frac{1}{\Gamma(\nu)2^{\nu-1}}\qty(\frac{\sqrt{2\nu}}{l}\norm{x_i-x_j}_2)^\nu K_\nu\qty(\frac{\sqrt{2\nu}}{l}\norm{x_i-x_j}_2)
    \label{eqn:Matern}
\end{equation}

where $l>0$ is a real-valued length scale parameter, $\Gamma(\cdot)$ is the Gamma function, $K_\nu(\cdot)$ is a modified Bessel function, and $\nu$ is a parameter determining the smoothness of the kernel \cite{scikit-learn}. They were combined with a Constant Kernel and a White Noise Kernel to account for variance in the mean and possible small amounts of noise respectively. In the results of Section 3
only the Mat\'{e}rn kernel was used with $\nu$ chosen to be $7/2$. The remaining kernel hyperparameters are tuned during fitting by maximizing the log-marginal-likelihood using the L-BFGS algorithm. The implementation is performed entirely using Python's \texttt{sklearn} library.

For a detailed presentation of Gaussian Process Regression see \cite{rasmussen2006cki}.

\section{Embedding Theorems}
\label{sec:appA}
The delay embedding theorems of Whitney and Takens guarantee that for any manifold $\mathcal{S}\subset\mathbb{R}^n$ parameterized by time, it is possible to construct a manifold $\mathcal{S}^*$ that is diffeomorphic to $\mathcal{S}$: there exists a differentiable bijective function $f:\mathcal{S}\to\mathcal{S}^*$ such that its inverse is differentiable as well. The reconstruction can be achieved using the time series of a one-dimensional observable that is coupled to all components of the dynamical system. An upper bound on the embedding dimension $k$ (number of delay observations of the one coordinate that is needed to construct $\mathcal{S}^*$) is $2n+1$ (i.e. $\mathcal{S}^*\subset\mathbb{R}^{2n+1}$).

These theorems are powerful and, in fact, suitable to address much more complicated problems, like chaotic dynamics. While in our problem we anticipate the dynamics to be much simpler, their appeal is their ability to uncover information about the system whose explicit dynamics are unknown, especially doing so through the use of a single observed variable (and its delays) as opposed to a larger number of observed quantities.

The Whitney embedding theorem \cite{whitney1936} can be restated as

\noindent
{\bf Whitney Embedding Theorem, 1936} 
Any $n$-dimensional, differentiable (of class $C^r$) manifold 
$M$ can be embedded in $R^{2 n +1}$ through a $C^r$ function. 

Takens \cite{takens1981} focused on dynamical systems for which 
an observable exists and  proved the following delay embedding 
theorem. Assume $M$ is an $n$-dimensional compact manifold. 
Let $X$ be a  vector
field on $M$ and $y:M \to \mathbb{R}$ be a smooth function 
(observable). Moreover, let $\phi$ denote the flow of $X$ on
$M$ (smooth diffeomorphism).

\noindent
{\bf Takens Delay Embedding Theorem, 1981} 
For pairs $(\phi, y)$, $\phi: M \to M$, it is a generic property
that the map $\Phi_{(\phi, y)}: M \to \mathbb{R}^{2n+1}$
defined by
\begin{equation}
\Phi_{(\phi, y)}(x) =(y(x), y(\phi(x)), ..., y(\phi^{2n}(x))) \end{equation}
is an embedding.

While the Takens theorem pertains to smooth attractors,
it was later generalized to apply to attractors
of any box-counting dimension \cite{sauer1991}. 
Another generalization is for more than one observables
\cite{deyle2011}.

\end{document}